\documentclass[prl,twocolumn,superscriptaddress,floatfix]{revtex4}
\usepackage{graphicx}
\begin{document}
\title{Experimental evidence of the vapor recoil mechanism in the boiling crisis}
\author{V. S. Nikolayev}\email[Web page: ]{http://www.pmmh.espci.fr/~vnikol}
\affiliation{ESEME, Service des Basses Temp\'eratures, DRFMC/DSM/CEA-Grenoble,\\
17 rue des Martyrs, 38054 Grenoble Cedex 9, France} \affiliation{CEA-ESEME, ESPCI-PMMH-P6-P7, 10, rue Vauquelin,
75231 Paris Cedex 5, France}
\author{D. Chatain}\affiliation{ESEME, Service des Basses Temp\'eratures, DRFMC/DSM/CEA-Grenoble,\\
17 rue des Martyrs, 38054 Grenoble Cedex 9, France}
\author{Y. Garrabos}
\affiliation{CNRS-ESEME, Institut de Chimie de la Mati\`{e}re Condens\'{e}e de Bordeaux,\\
Universit\'{e} de Bordeaux I, Avenue du Dr. Schweitzer, 33608 Pessac Cedex, France}
\author{D. Beysens}
\affiliation{ESEME, Service des Basses Temp\'eratures, DRFMC/DSM/CEA-Grenoble,\\
17 rue des Martyrs, 38054 Grenoble Cedex 9, France} \affiliation{CEA-ESEME, ESPCI-PMMH, 10, rue Vauquelin, 75231
Paris Cedex 5, France}
\date{\today} 

\begin{abstract}
Boiling crisis experiments are carried out in the vicinity of the liquid-gas critical point of H$_2$. A magnetic
gravity compensation set-up is used to enable nucleate boiling at near critical pressure. The measurements of
the critical heat flux that defines the threshold for the boiling crisis are carried out as a function of the
distance from the critical point. The obtained power law behavior and the boiling crisis dynamics agree with the
predictions of the vapor recoil mechanism and disagree with the classical vapor column mechanism.
\end{abstract}\maketitle

Boiling is a highly efficient way to transfer heat. This is why it is widely used, in particular in high power
industrial heat exchangers, e.g. nuclear power plant steam generators. Boiling is often considered by the
physics community as a well understood phenomenon, at least qualitatively. It is true that boiling has been
studied extensively by experiment for common fluids and conventional regimes, for instance for water at
atmospheric pressure and moderate heat flux supplied to the fluid. However, the basic theory of boiling remains
\emph{terra incognita}, in particular, the phenomena very close to the heating surface, at a scale much smaller
than the vapor bubbles \cite{Dhir}.

The efficiency of industrial heat exchangers increases with the heat flux. However, there is a limit called
Critical Heat Flux (CHF). It corresponds to a transition from nucleate boiling (boiling in its usual sense) to
film boiling where the heater is covered by a quasi-continuous vapor film and the evaporation occurs at the
gas-liquid interface. Since the gas conducts heat much less than the liquid, the heat transfer efficiency drops
sharply during this transition and the heater heats up, which may cause its damage if the power is not cut
immediately. This transition is called ``burnout", ``departure from nucleate boiling" or ``Boiling Crisis" (BC).

Among several dozens of existing models of BC, the Zuber approach \cite{Dhir,Theo} is the only one that can be
considered as a theory, the others being mainly empirical. According to this model, vapor columns form at the
nucleation sites on the heater. The vapor moves upwards while the liquid moves to the bottom of the column where
evaporation occurs. This counter-flow motion induces the Kelvin-Helmholtz instability, which leads to the
destabilization of the whole system and to the creation of a vapor film on the heater. The transition occurs
when the vapor velocity exceeds a threshold \cite{Theo} resulting in the following CHF expression,
\begin{equation}\label{Zub}
   q_{CHF}\sim H[\sigma g (\rho_L-\rho_V)\rho_V^2]^{1/4},
\end{equation}
where $H$ is the latent heat, $\rho_L$ ($\rho_V$) is the density of the liquid (gas) phase, $\sigma$ is the
surface tension, and $g$ is the gravity acceleration. While this expression fits a number of experimental data
sets, the underlying physics is questionable. Indeed, the vapor column morphology of boiling is quite rarely
observed while the BC exist for almost all morphologies of boiling, for pool boiling (i.e., natural convection
boiling) or for flow boiling (i.e. boiling of the fluid flowing in a heated tube). Besides, many experimental
results, in particular those obtained in low-gravity \cite{Straub}, cannot be fitted by Eq.~(\ref{Zub}). Other
physical phenomena should then be responsible for the triggering of BC. A strong dependence of CHF on the
wetting properties of the heater \cite{Dhir} suggests a phenomenon at the contact line level.

\begin{figure}[htb]
\centering
\includegraphics[width=6cm]{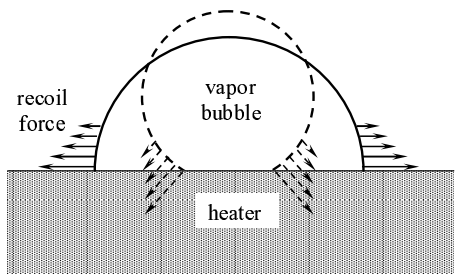}
\caption{Sketch illustrating how the vapor recoil initiates the bubble spreading. The amplitude and direction of
the vapor recoil force are shown by arrows.} \label{brec}
\end{figure}

A vapor recoil mechanism for BC has already been proposed in \cite{EuLet99,IJHMT01}. A fluid molecule leaving
the liquid interface causes a recoil force analogous to that created by the gas emitted by a rocket engine. It
pushes the interface towards the liquid side in the normal direction. An average vapor recoil force appears
because the fluid necessarily expands while transforming from the liquid to the gas phase. The stronger the mass
evaporation rate $\eta$ (per time and interface area), the larger the vapor recoil force. One finds that the
vapor recoil force per interface area is $P_r=\eta^2(\rho_V^{-1}-\rho_L^{-1})$ \cite{EuLet99}. The evaporation
is particularly strong in the vicinity of the contact line of a bubble, inside the superheated layer of the
liquid (Fig.~\ref{brec}).
The resulting vapor recoil force can pull apart the bubble contact line and make it spread over the heater, thus
creating a nucleus for the vapor film. The results of the vapor recoil model are in qualitative agreement with
the observations by some of us \cite{PRE01} and those by other groups \cite{Kandlikar,Nishio}. The present
Letter deals with its quantitative verification.

Consider the fluid far from its critical point, where the system pressure $p\ll p_c$ ($p_c$ is the critical
pressure). Instead of using pressure, it is more convenient to argue in terms of the saturation temperature $T$
from which $p$ can be obtained readily. An estimation \cite{EuLet99} shows that the heat flux necessary to
create a vapor recoil force comparable to that of the surface tension corresponds to the experimental CHF order
of magnitude. If the contributions of the vapor recoil and the surface tension $\sigma$ are of the same order,
\begin{equation}\label{rat}
    \frac{P_rl_c}{\sigma}\sim\frac{q_{CHF}^2l_c}{H^2\sigma}\left(\frac{1}{\rho_V}-\frac{1}{\rho_L}\right)=\mbox{const},
\end{equation}
where the capillary length $l_c=\sqrt{\sigma/g(\rho_L-\rho_V)}$ is the natural lengthscale. Eq.~(\ref{rat})
results in a CHF expression identical to Zuber's expression (\ref{Zub}) if the inequality $\rho_L\gg\rho_V$ is
taken into account. Both models are then difficult to distinguish far from the critical temperature $T_c$.

On the contrary, their behaviors close to $T_c$ are quite different. A reasoning presented in \cite{PRE01} led
to the following vapor recoil model result
\begin{equation}\label{tchf}
  q_{CHF}\sim(T_c-T)^{1+\nu-3\beta/2},
\end{equation}
where $\beta=0.325$ and $\nu=0.63$ are the universal critical exponents, $1+\nu-3\beta/2=1.14$. The Zuber
expression (\ref{Zub}) also leads to a power law with exponent $5\beta/4+\nu/2=0.72$. This value can be obtained
from the scaling relations $H\sim\rho_L-\rho_V\sim(T_c-T)^\beta$ and $\sigma\sim(T_c-T)^{2\nu}$.

The thermal diffusivity vanishes at $T_c$. The thermally controlled bubble dynamics is thus slower than at low
pressure (critical slowing down) and the CHF is much smaller. Optical distortions inevitable at low pressures
because of violent fluid motion and strong temperature gradients \cite{Kenning} are nonexistent at $T\simeq T_c$
where very detailed observations can thus be performed. However, the surface tension becomes very low and
gravity flattens the gas-liquid interface. Reduced gravity conditions are thus necessary to preserve the
existence of bubbles, hence the nucleate boiling itself.

\begin{figure}[htb]
\centering
\includegraphics[width=\columnwidth]{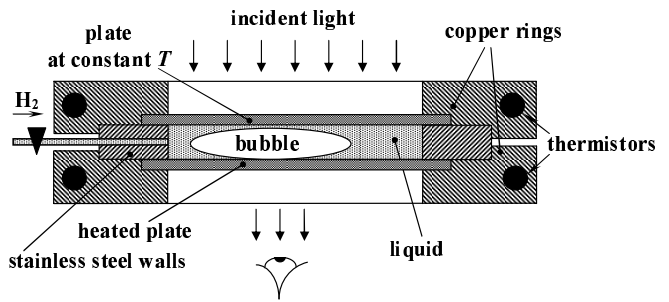}
\caption{Sketch of the cylindrical transparent experimental cell.} \label{cell}
\end{figure}

The cryogenic magnetic levitation installation at CEA-Grenoble \cite{Cryo} was used to achieve a gravity
compensation. The accuracy is of 2\% in the cylindrical fluid volume of 8~mm diameter and 5~mm thickness. The
experimental cell (Fig.~\ref{cell})
can be partially filled or pumped off \emph{in situ} by using a capillary equipped with a cryogenic electric
valve. The latter remains closed during the experiment. The cell is filled with H$_2$ at critical density
$\rho_c$ so that the gas phase occupies a half of the cell independently of $T$ because of the symmetry of the
co-existence curve with respect to $\rho_c$, $(\rho_L+\rho_V)/2=\rho_c$. During the evolution, the bubble mass
can vary while its volume does not change.

The best observations of BC \cite{Theo,Kandlikar,Nishio,Van,Tor,Jamet} involved a transparent heater to detect
and follow the heater dryout. Our cylindrical cell (Fig.~\ref{cell}) has transparent sapphire end plates.
Several thermistors are integrated into copper rings that have a good thermal contact with the plates. Both
rings are connected thermally to a colder liquid helium bath with wires that serve as thermal resistances. The
thermistors are used to inject the controlled heat power into one of the plates that serves as a heater and to
maintain the temperature of the other plate with 1~mK precision (in a stationary state) by a temperature
regulation system. Sapphire is an excellent heat conductor in the cryogenic temperature range ($T_c=33$K for
H$_2$). The lateral cell wall is made of stainless steel, with conductivity about $10^3$ times less than that of
the sapphire.

Because of the complete wetting conditions characteristic of near critical fluids \cite{PRE01}, the wetting
layer always covers the cell at equilibrium. Due to this fact, a good thermal contact of the temperature
controlled plate with the rest of the fluid is provided. The cell location with respect to the magnetic field is
chosen in such a way that a residual magnetic force (which plays the role of an effective gravity) positions the
bubble against the heating plate. This effective gravity field is directed to the cell center (\cite{Cryo}, see
also the sketch in Fig.~\ref{dry_spot}a below) so that the denser liquid phase is attracted to the cell center.
Unfortunately, it is quite difficult to quantify the force acting on the bubble since there is no possibility to
map the magnetic field with a sufficient precision.

The surface tension prevents the liquid from gathering in the cell center by keeping the bubble convex, and the
bubble image is circular far from $T_c$. Close to the critical point, the surface tension becomes too weak and
the liquid gathers in the center. Since the wetting layer remains at the cell walls, the bubble takes an unusual
torus shape, see Fig.~\ref{dry_spot}. This occurs at $T=T_g\approx 32.9$K. We call this geometry annular because
of the observed bubble image.

For each experimental run, a thermal regulation reference temperature is chosen and defines $T$. By tuning the
heater power, the cell is thermally equilibrated. Then heating is increased as needed and a stationary boiling
state is obtained after a transition period. The heater power $P$ and the heater temperature $T_h$ are then
recorded.

\begin{figure}[htb]
\centering
\includegraphics[width=6cm]{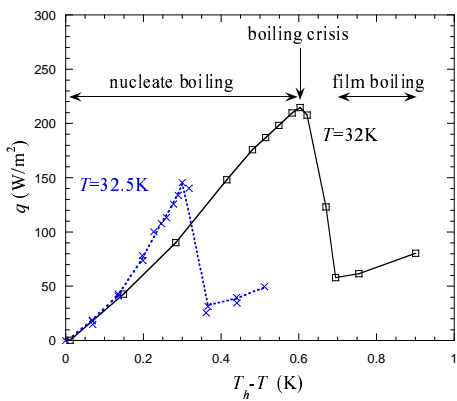}
\caption{Examples of the boiling curves for different pressures corresponding to two indicated values of $T$.
The boiling regimes are indicated (32K curve). The lines are guides for the eye.} \label{boilcurve}
\end{figure}

$T_h$ has been measured also with the empty cell for different $T$ and heating powers $P_e$ and the dependence
$P_e=P_e(T_h,T)$ was established. The amount of heat transferred across the cell walls of the filled cell can
thus be determined by calculating this function for the values of both temperatures measured in the filled cell.
The heat flux $q$ carried by the fluid is obtained by dividing $P-P_e$ by the heater area. The dependence of $q$
(calculated as explained above) on $T_h$ is called the boiling curve (Fig.~\ref{boilcurve}).

\begin{figure}[htb]
\centering
\includegraphics[width=\columnwidth]{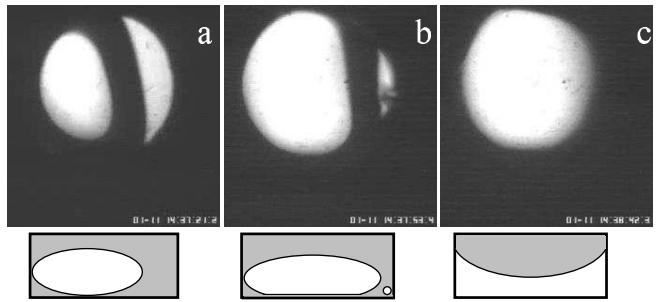}
\caption{Heater drying dynamics at the CHF at $T=32.5$K visualized through the transparent heater (closest to
the observer). (a) Initial bubble position at equilibrium. (b) Bubble partially spread. (c) The heater is
completely dried out. The section of the cell interior with a plane perpendicular to the image is sketched below
the corresponding photo. On the sketches, the vapor is white and the liquid is gray; the heater is at the
bottom.} \label{circular}
\end{figure}

\begin{figure}[htb]
\centering
\includegraphics[width=\columnwidth]{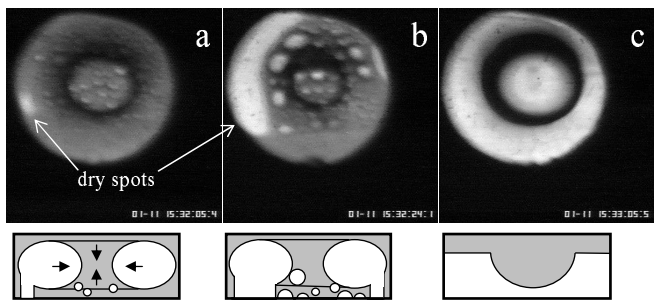}
\caption{Heater drying dynamics slightly above the CHF at $T=32.95$K (in annular geometry, the large bubble has
a toroidal shape as described in the text). The bright areas on the photos are the dry spots. Similarly to
Fig.~\ref{circular}, phase distributions for each photo are shown in the sketches. (a) Beginning of the dry spot
(small white spot to the left) growth. (b) Intermediate stage. (c) Complete drying of the heater; the liquid
phase has taken a shape of a hat seen from the top. Nucleated small bubbles are visible in (a-b). The direction
and the relative magnitude of the effective gravity are shown by arrows in the sketch (a).} \label{dry_spot}
\end{figure}

The cell can be observed optically through the plates by using a light source, a CCD camera and two periscopes.
Far from $T_c$, the bubble is circular (Fig.~\ref{circular}).
The nucleation, growth and departure of small gas bubbles occur at the periphery of the heating plate where the
liquid wetting layer is thicker. Since the large bubble occupies most of the plate image and its curved surface
looks dark, the small bubbles are almost invisible. Their presence can be detected by the trembling of the large
bubble during their coalescence. A sudden disappearance of the trembling indicates the complete dryout of the
liquid film (Fig.~\ref{circular}c). The nucleation of the dry spot and the contact line motion is difficult to
observe in this geometry without special optical means \cite{John}. At $T\lesssim T_g$ the wetting layer
thickens and the contact line becomes visible in motion but the optical contrast is low.

At the CHF the BC does not occur immediately after the temperature rise. The cell first attains a nearly
stationary state where $T_h$ fluctuates slightly. One of the fluctuations then leads to a rapid drying of the
major part of the heater. At the CHF, the liquid loses completely its contact with the heater which corresponds
to film boiling. The transferred heat flux $q$ falls sharply (Fig.~\ref{boilcurve}). Practically no fluid motion
is observed any more even when the heating power is increased.

The optical contrast is much better in the annular regime ($T_g<T<T_c$) because the wetting layer is several
times thicker. Its thickness can be judged from the maximum size of small bubbles that nucleate and grow inside
it (Figs.~\ref{dry_spot}a,b).
Small bubbles also form in the plate center, which is convenient for their observation. They grow and slide to
the plate periphery or depart from the plate under the action of the effective gravity, which pushes them in the
direction of the large bubble. They eventually coalesce with it.

At $q\lesssim q_{CHF}$ dry spots under the small bubbles begin to appear and disappear intermittently when the
bubbles depart from the heater. A bubble that appears in the hottest point (where the liquid layer is thinner,
presumably due to the vapor recoil pressure) coalesces with the toroidal bubble and forms a dry spot which
remains stationary. This bubble forms intermittently a ``bridge" connecting the large bubble to the heater as
sketched in Fig.~\ref{dry_spot}a. At $q=q_{CHF}$ a large dry spot also appears (Fig.~\ref{dry_spot}a) and keeps
growing (Fig.~\ref{dry_spot}b). The smaller dry spots under other bubbles keep appearing and disappearing but
grow larger. All the dry spots grow so large that they coalesce and, suddenly, the heater dries out completely
(Fig.~\ref{dry_spot}c). This picture is in full analogy with the observations by other groups
\cite{Nishio,Kandlikar} performed at much lower pressures. As expected, the BC slows down near $T_c$, and can
take as long as 1-2 min for the closest to $T_c$ runs.

The thickness of the wetting layer reflects the force that presses the bubble against the heater. When the layer
is thicker (smaller force), a larger heat flux is needed to dry out the heater and the CHF is larger. This is
what happens after a cell displacement with respect to the magnetic field or after a bubble topology change. The
$q_{CHF}(T)$ dependence should thus be measured at the same (circular) bubble topology, i.e. at $T<T_g$.

\begin{figure}
  \begin{center}
  \includegraphics[width=6cm]{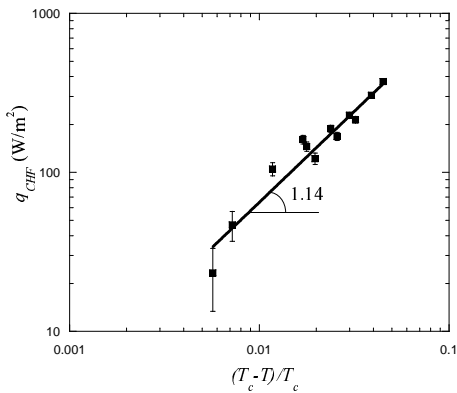}\\
  \end{center}
\caption{The critical heat flux as a function of the distance to the critical point. The solid straight line is
the vapor recoil model prediction Eq.~(\ref{tchf}).}\label{chf-t}
\end{figure}

The $q_{CHF}(T)$  dependence is shown in Fig.~\ref{chf-t} and compared with the vapor recoil model prediction,
Eq.~(\ref{tchf}).
A good agreement is found, which demonstrates the validity of the model. It is evident that the data cannot be
fitted well with the Zuber equation (\ref{Zub}) as the corresponding slope is nearly twice smaller.

The boiling crisis has been observed at high, nearly critical pressure and at low gravity. At these conditions
the BC is triggered by the growth of dry spots under individual vapor bubbles and is qualitatively analogous to
the BC at normal gravity and low pressures. The dry spot growth is followed by the bubble coalescence provoking
heater dryout. At low pressures, the vapor recoil model gives a CHF expression similar to the classical Zuber
formula. At high pressures, the two expressions, however, differ strongly. The measurements of the CHF depending
on the distance to the critical point demonstrate the validity of the vapor recoil model.

These results open the way to more precise numerical simulation that can now be based on a well identified
physical phenomenon. The CHF can then be predicted as a function of various system parameters such as pressure,
material properties, geometry, gravity level, etc.

\begin{acknowledgments}
This work is partially supported by CNES. We thank D.~Communal for his contribution to the thermal regulation
system and P. Seyfert for his helpful comments.
\end{acknowledgments}

\end{document}